\documentclass{iau}
\usepackage{graphicx}
\usepackage{natbib}

\title[Sgr A* observations with VERA and KaVA]{Long-term monitoring of Sgr A* at 7 mm with VERA and KaVA}

\author[K. Akiyama et al.]{K. Akiyama$^{1,2}$, M. Kino$^{3}$, B. Sohn$^{5}$, S. Lee$^5$, S. Trippe$^{6}$, M. Honma$^{2,4}$, KaVA AGN WG et al.}

\affiliation{
$^1$Department of Astronomy, Graduate School of Science, The University of Tokyo \\ email: {\tt kazunori.akiyama@nao.ac.jp} \\[\affilskip]
$^2$Mizusawa VLBI Observatory, National Astronomical Observatory of Japan \\[\affilskip]
$^3$Institute of Space and Astronautical Science, Japan Aerospace Exploration Agency \\[\affilskip]
$^4$Graduate University for Advanced Studies \\[\affilskip]
$^5$Korea Astronomy \& Space Science Institute \\[\affilskip]
$^6$Department of Physics and Astronomy, Seoul National University
}

\pubyear{2013}
\volume{303}  
\pagerange{xxx--xxx}
\date{2013/11/20 and in revised form ??}
\jname{The Galactic Center: Feeding and Feedback in a Normal Galactic Nucleus}
\editors{A.C. Editor, B.D. Editor \& C.E. Editor, eds.}

\newcommand{\Mdot}{\dot{M}}

\newcommand{\araa}{ARAA}
\newcommand{\apj}{ApJ}
\newcommand{\apjl}{ApJL}

\newcommand{\nat}{Nature}
\newcommand{\aap}{A\&A}
\newcommand{\mnras}{MNRAS}
\newcommand{\aj}{AJ}
\newcommand{\pasj}{PASJ}
\newcommand{\pasp}{PASP}
\newcommand{\jcap}{JCAP}

\begin{document}

\maketitle
\begin{abstract}
We present the results of radio monitoring observations of Sgr A* at 7 mm (i.e. 43 GHz) with VLBI Exploration of Radio Astrometry (VERA), which is a VLBI array in Japan. VERA provides angular resolutions on millisecond scales, resolving structure within ~100 Schwarzschild radii of Sgr A* similar to Very Large Baseline Array (VLBA). We performed multi-epoch observations of Sgr A* in 2005 - 2008, and started monitoring it again with VERA from January 2013 for tracing the current G2 encounter event. Our preliminary results in 2013 show that Sgr A* on mas scales has been in ordinary state as of August 2013, although some fraction of the G2 cloud already passed pericenter of Sgr A* in April 2013. We will continue on monitoring Sgr A* with VERA and newly developed KaVA (KVN and VERA Array).
\end{abstract}

\section{Introduction}

There are plenty of evidences for the presence of a super-massive black hole (SMBH) with a mass of $4\times10^{6}\,{\rm M}_{\odot}$ \citep[e.g.,][]{Gillessen2009} associated with the Galactic center Sagittarius A* (Sgr A*). The angular size of its Schwarzschild radius ($1\, {\rm R_{{\rm Sch}}} \sim 10\,{\rm \mu as}$) is the largest among known black hole candidates $ $due to its proximity \citep[$D\sim 8$ kpc, e.g.,][]{Gillessen2009,Honma2012}, providing an unprecedented opportunity to probe the environment around the SMBH.

The Very Long Baseline Interferometer (VLBI) can achive a spatial resolution within 100 ${\rm R_{sch}}$. Though detailed imaging of its structure has suffered from the effect of interstellar scattering \citep[e.g.,][]{Lo1998}, its intrinsic structure start appearing at wavelength shorter than {$\sim$}1 cm \citep[e.g.,][]{Bower2004}. The intrinsic size of Sgr A* measured by mm-VLBI observations has a wavelength-dependence \citep[e.g.][]{Falcke2009}, interpreted to originate in stratified wavelength-dependent photosphere \citep[e.g.][]{Loeb2007,Falcke2009}. Furthermore, variations of its intrinsic size are detected at 7 mm \citep[][]{Bower2004}

The flux of Sgr A* is known to be variable at radio, NIR and X-ray bands on various time-scales from intra-day to a few months. The radio band variability on time scales longer than a day could be explained by both an intrinsic origin and interstellar scintillation. However, recent studies show that the flux variability at short cm/mm wavelengths cannot be explained by a simple model of interstellar scattering but is likely to be intrinsic \citep{Herrnstein2004,Lu2011}.

Monitoring of Sgr A* with VLBI is of great importance to investigate relation between variations in the radio flux and size at the vicinity of its SMBH, and probe a possible mechanism of its variability. To do this, We performed multi-epoch observations of Sgr A* with VLBI Exploration of Radio Astrometry (VERA).

\section{VERA Observations in 2005 - 2008\label{sec:obs_2005-2008}}

\begin{figure}
\begin{center}
\includegraphics[width=0.5\hsize]{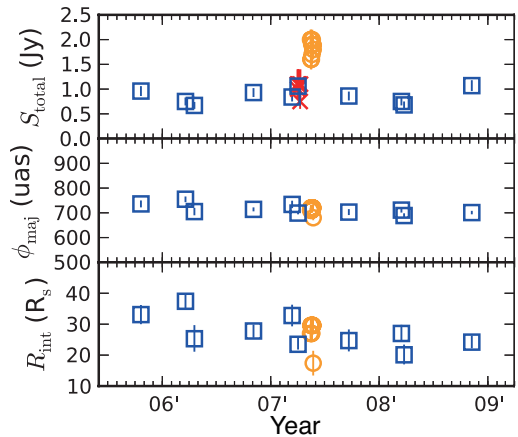}
\caption{The time variations of the flux and structure of Sgr A* measured with VERA in 2005-2008 \citep{Akiyama2013}. From top to bottom, vertical axes correspond to flux, size of the major axis, and intrinsic size, respectively. The squares are our results, while the circles and the crosses are VLBA results. \label{fig:KA2013}}
\end{center}
\end{figure}

We performed multi-epoch observations of Sgr A* at 7 mm (i.e. 43 GHz) from 2004 to 2008 \citep{Akiyama2013}. We detected quiescent flux levels and no radio flares (see, Figure \ref{fig:KA2013}). However, comparisons with previous Very Long Baseline Array (VLBA) results showed that Sgr A* underwent a flaring event for at least 10 days in 2007 May reported in \citet{Lu2011}. Our data show that the intrinsic size of Sgr A* remained (within errors) unchanged compared to the size before and after the flaring event, indicating that the brightness temperature of Sgr A* was increased. The duration of this flaring event is less than 31 days, which is shorter than the refractive time scale of 3 months \citep{Bower2004}. Moreover, it is difficult to explain the increase in the spectral index at this flaring event reported in \citet{Lu2011} by a simple interstellar scattering model \citep{Rickett1990}. Hence, the flaring event is most likely associated with changes in the intrinsic properties of Sgr A*.

Considering observed structure, it is unlikely that the flaring event is associated with an ejection of relativistic component or a temporal one-shot plasma heating such as an expanding plasma blob \citep[e.g.][]{Yusef2006} or a hot spot orbiting around the central black hole \citep[e.g.][]{Broderick2006}. Thus, the flaring event is likely to be associated with a brightness increase of the photosphere. Since the synchrotron cooling time-scale at mm wavelengths is much shorter than than the duration of the flare, understanding the flare requires a mechanism that heats electrons continuously on timescales much longer than the orbital timescales of the accretion disk such as a standing shock in an accretion flow. In the future, simultaneous multi-frequency flux monitoring from cm to sub-mm wavelengths would be helpful to constrain the properties of electron distribution at the flaring event.

\section{New VERA Observations from January 2013\label{sec:obs_2013}}
Recently, the gas cloud G2 with three earth masses has been detected on its way to Sgr A* \citep[][]{Gillessen2012}. G2 will approach pericenter at a distance of only 1500-2200 ${\rm R_{sch}}$ (15-22 mas) in early-2014 \citep[][]{Phifer2013,Gillessen2013}. The dynamic evolution around Sgr A* in the next few years will be an unique probe for understanding the properties of the accretion flow/jet and the feeding processes of the SMBH.

Related with the G2 encounter with Sgr A*, two flaring phenomena are expected in radio regime as of November 2013. 
First, G2 is expected to trigger flaring of Sgr A*, since the radio flux is strongly related to the mass accretion rate $\Mdot$ to the central black hole, scaling with $\Mdot ^2$ \citep[e.g.][]{Mahadevan1997} for the accretion flow and $\Mdot ^{1.4}$ \citep[e.g.][]{Falcke1995} for the jet. A possible increase in the mass accretion rate would cause a flux increase as well as structural changes in the visible structure detectable with VLBI, like an increase in the size of the radio emitting region \citep[e.g.][]{Moscibrodzka2012} or appearance of the visible jet. Second, \citet{Narayan2012} predict that the interaction between G2 and the accretion disk will produce a bow-shock luminous in radio regime around pericenter. VLBI observations would be important for constraining the size and the compact feature of the shock region.

\begin{figure}
\centering{}%
\begin{tabular}{cc}
\begin{minipage}[t]{0.52\columnwidth}%
\begin{center}
\includegraphics[width=1\textwidth]{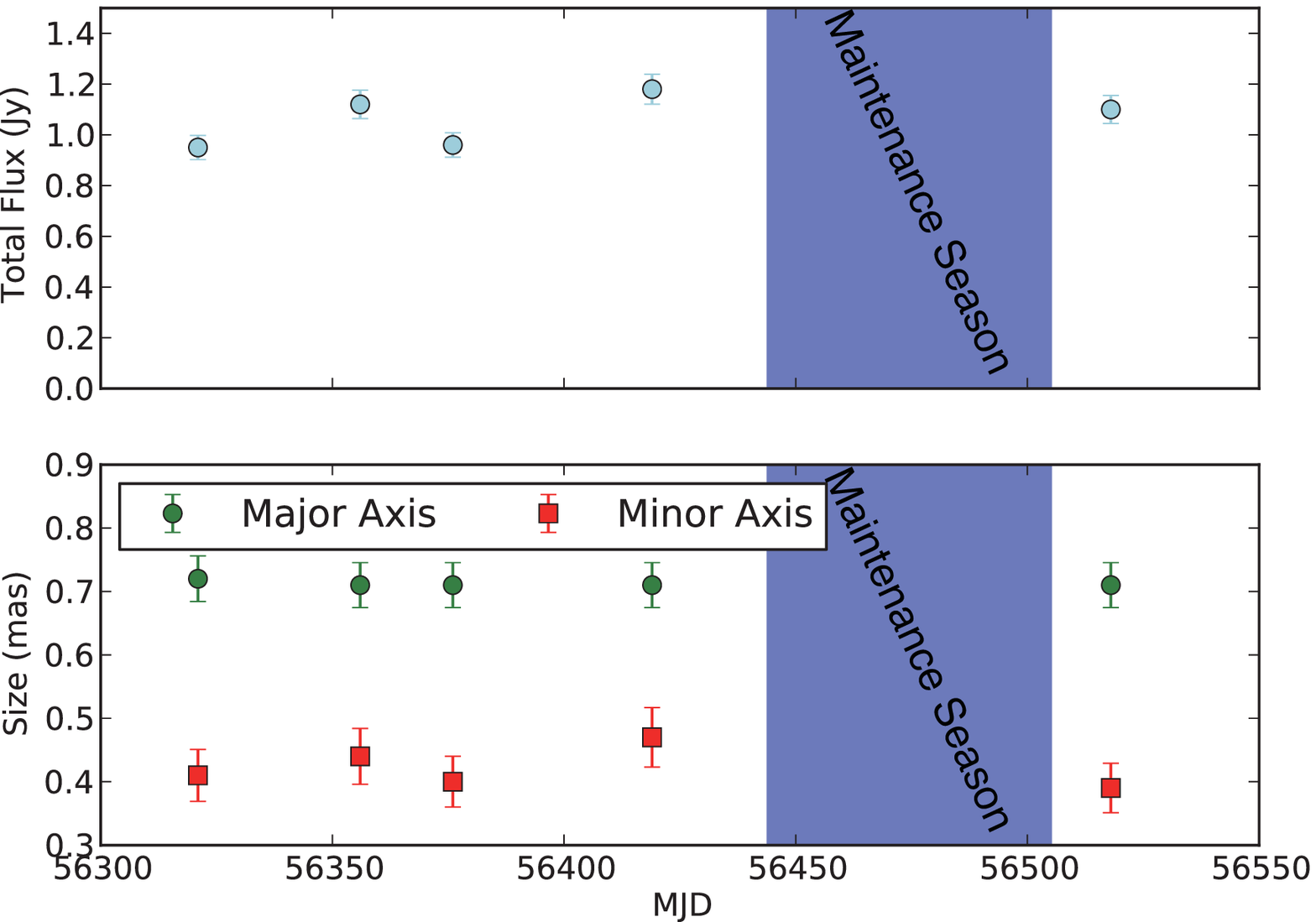}
\par\end{center}%
\end{minipage} & %
\begin{minipage}[t]{0.36\columnwidth}%
\begin{center}
\includegraphics[width=1\textwidth]{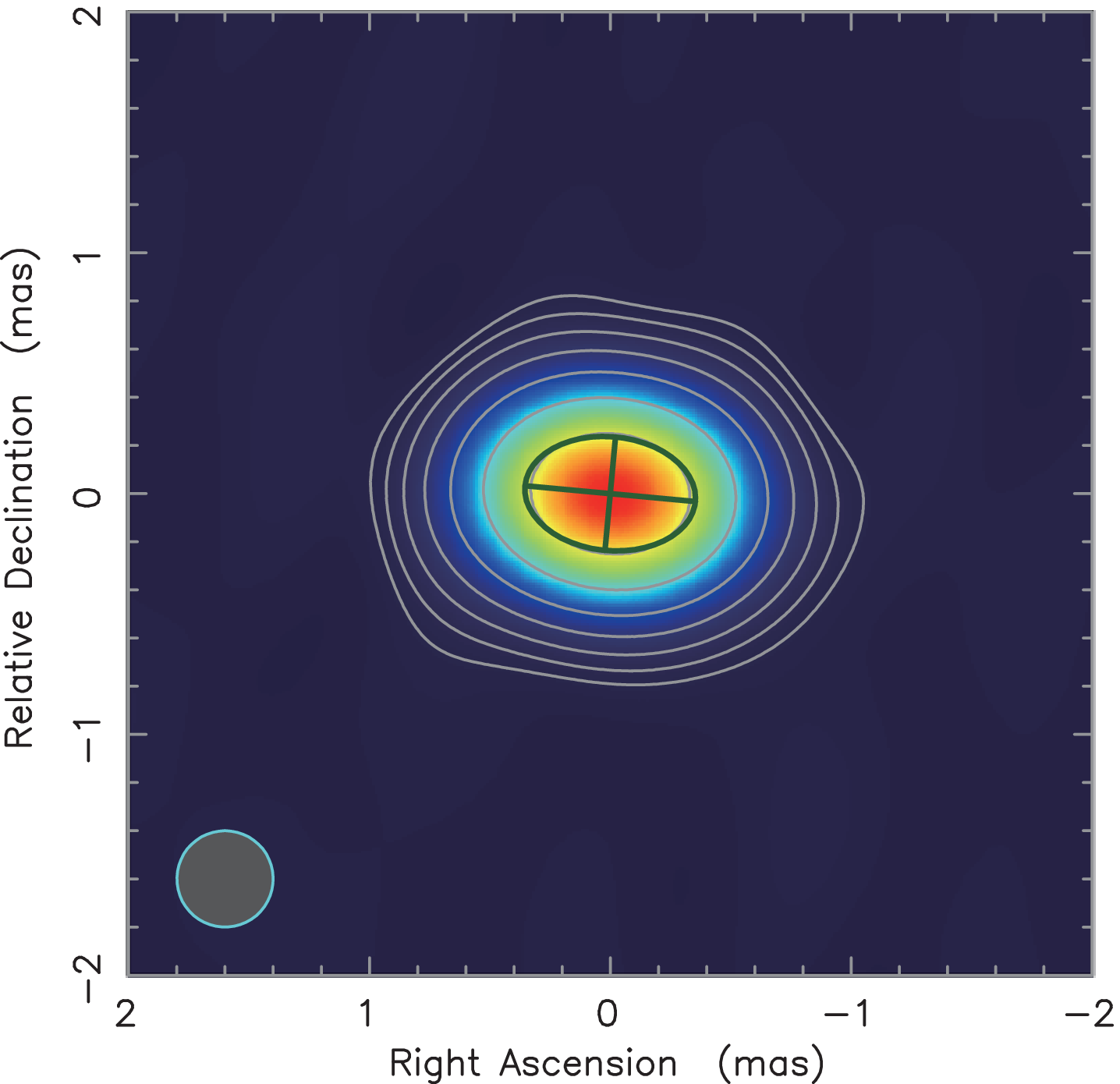}
\par\end{center}%
\end{minipage}\tabularnewline
\end{tabular}
\caption{Preliminary results of VERA observations at 7 mm (i.e. 43 GHz) in the first half of 2013. Left: The upper panel shows a time variation in the total flux, while the lower panel shows a time variation in the major/minor
axis size of the best-fit elliptical Gaussian model. Right: The image at 7 mm on 7th May 2013, with the best-fit elliptical Gaussian model (shown in a green line). The image is restored using a circular Gaussian beam with a FWHM of 0.4 mas.}
\end{figure}

In order to trace this exciting event, we performed monitoring observations with VERA at 7 mm with an interval of 3 weeks. Here, we show preliminary results by August 2013 in Figure \ref{sec:obs_2013}. The radio flux and the major/minor axis size show a variation within 10 \% level, consistent with previous activities. Our preliminary results show that Sgr A* on mas scales has been quiescent as of August 2013, although the head part of G2 already passed pericenter in April 2013 \citep{Gillessen2013}. Considering a free-fall time scale of few months from pericenter to the central SMBH, our results indicate that the feeding rate of the G2 head part is comparable or less than the current accretion rate of Sgr A*, or most of the accreted gas is lost on the way to the black hole via a disk outflow in the accretion disk.

\section{Future observations with VERA and KaVA\label{sec:future}}

\begin{figure}
\centering{}%
\begin{tabular}{ccc}
\begin{minipage}[t]{0.35\columnwidth}%
\begin{center}
\includegraphics[width=1\textwidth]{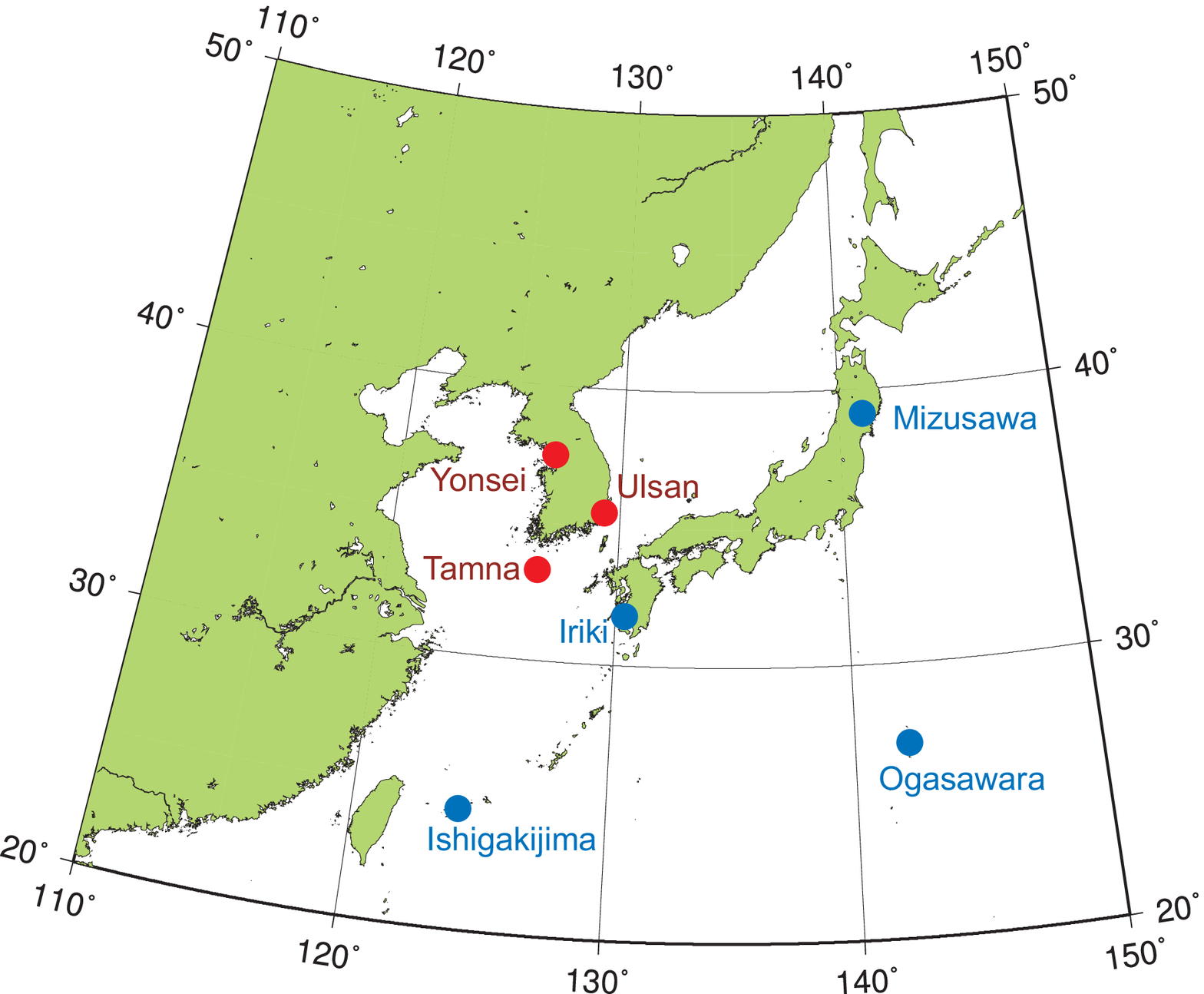}
\par\end{center}%
\end{minipage} & %
\begin{minipage}[t]{0.3\columnwidth}%
\begin{center}
\includegraphics[width=1\textwidth]{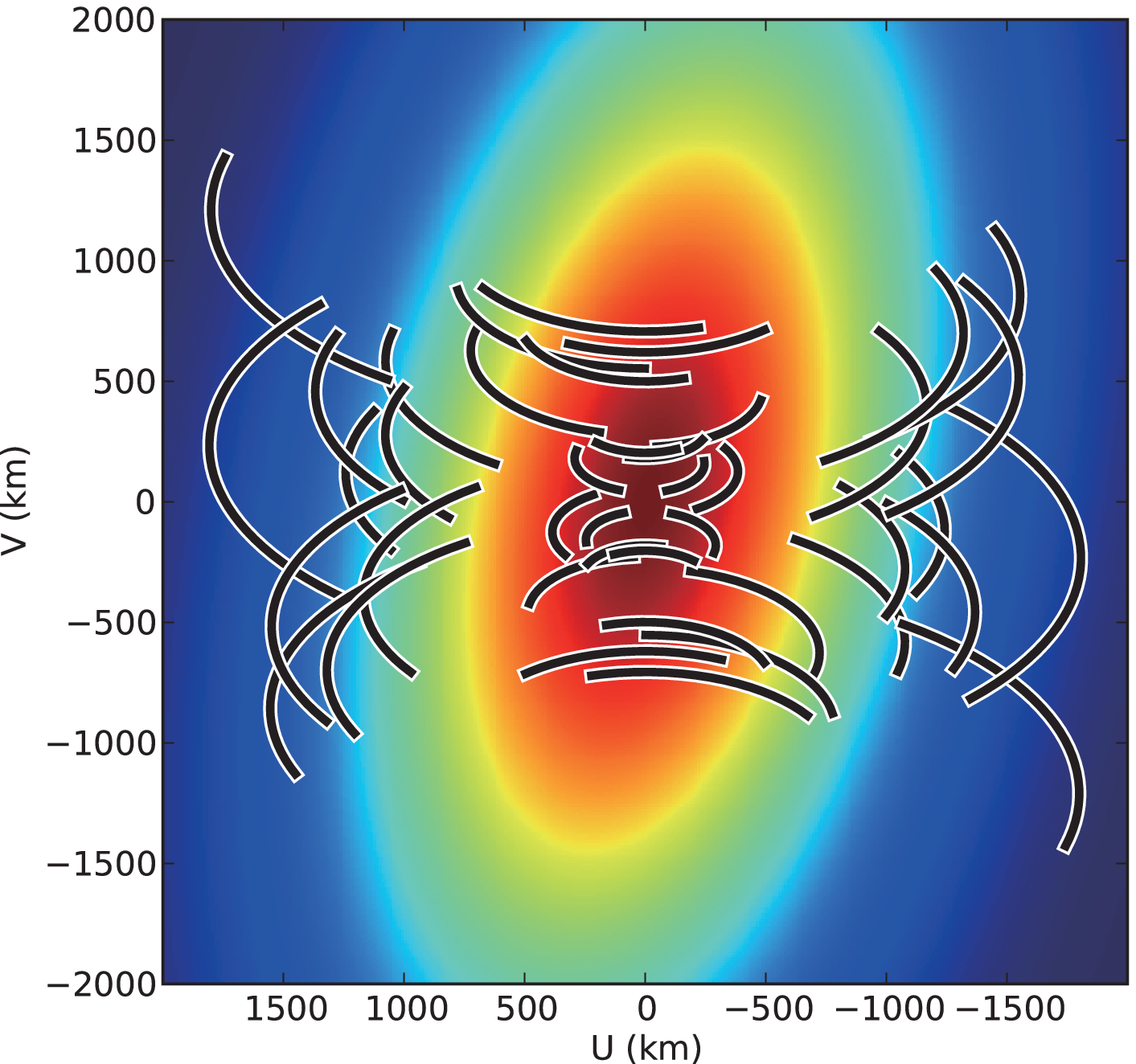}
\par\end{center}%
\end{minipage} & %
\begin{minipage}[t]{0.3\columnwidth}%
\begin{center}
\includegraphics[width=1\textwidth]{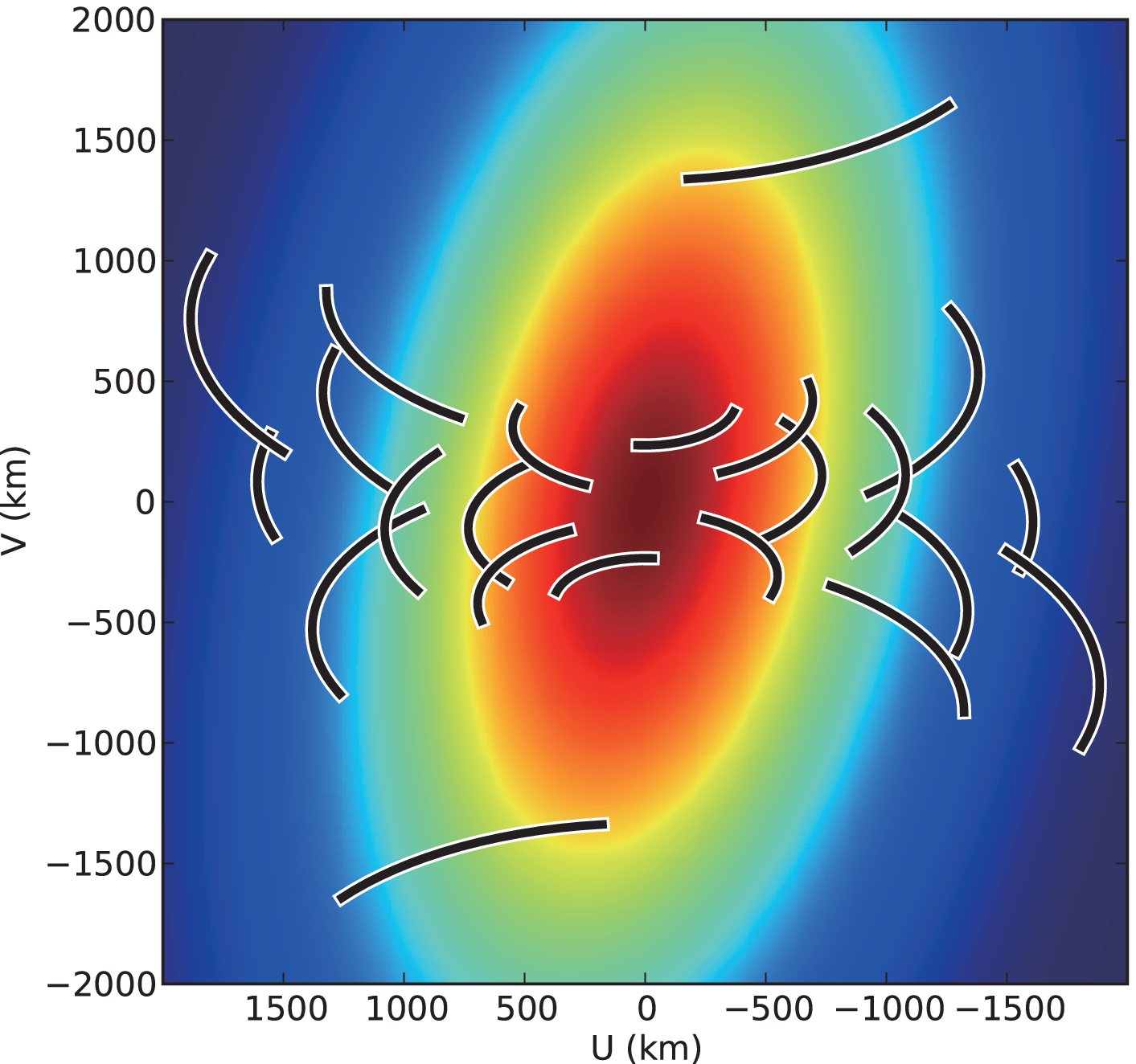}
\par\end{center}%
\end{minipage}\tabularnewline
\end{tabular}\caption{Left panel: Array configuration of KaVA. KaVA is a combination of Korean VLBI Network (KVN; colored in red) and VLBI Exploration of Radio Astrometry
(VERA; colored in blue). Middle and right panels: The $uv$-coverage of KaVA (the middle panel) and VLBA (the right panel) for Sgr A*. Lines are the $uv$-coverage of each array. The colored contour is the visibility amplitude distribution of the best-fit elliptical Gaussian model in \citet{Bower2004}. \label{fig:KaVA}}
\end{figure}

We are going to continue monitoring with newly developed KVN and VERA Array (KaVA). KaVA consists of 7 stations in Korean VLBI Network \citep[KVN,][]{Lee2011} and VERA (see, Figure \ref{fig:KaVA}), and will start regular observations at 13/7 mm (i.e. 22/43 GHz) from March 2014.

KaVA is expected to achieve a good performance for Sgr A* observations, since it has more short baselines than other VLBI arrays. One of reasons making it difficult to calibrate VLBI data of Sgr A* is its low correlated flux density in long baselines owing to the effect of the interstellar scattering. Short ($\leq \sim 2000$ km) baselines in KaVA provide more effective sampling of the visibilities of Sgr A* than VLBA (Figure \ref{fig:KaVA}), enabling accurate determination of its size and good-quality images similar to or better than VLBA. Our future observations will provide more detailed information about mas-scale structure of Sgr A* around the most important period of the G2 encounter.

\end{document}